\documentclass{elsart}
\usepackage{amsmath,amssymb}
\usepackage{cite}
\usepackage{graphicx}
\usepackage{epsfig}
\usepackage{subfigure}

\begin{document}

\begin{frontmatter}

\begin{flushright}
YITP-SB-12-40 \\ ICCUB-12-356
\end{flushright}

\title{Dark Radiation Confronting LHC in Z' Models}
\author[ub]{A. Solaguren-Beascoa},
\author[icrea,ub,cnyitp]{M.~C.~Gonzalez-Garcia}
\address[icrea]{Instituci\'o Catalana de Recerca i Estudis Avan\c{c}ats 
(ICREA)}
\address[ub]{ 
Departament d'Estructura i Constituents de la Mat\`eria and ICC-UB\\
  Universitat de Barcelona, Diagonal 647, E-08028 Barcelona, Spain.}
\address[cnyitp]{C.N. Yang Institute for Theoretical Physics\\
  State University of New York at Stony Brook\\
  Stony Brook, NY 11794-3840, USA.}
\begin{abstract}
Recent cosmological data favour additional relativistic degrees of
freedom beyond the three active neutrinos and photons, often referred
to as ``dark radiation''. Extensions of the SM involving TeV-scale
$Z'$ gauge bosons generically contain superweakly interacting light
right-handed neutrinos which can constitute this dark radiation.  
In this letter we confront the requirement on the parameters of the $E_6$
$Z'$ models to account for the present evidence of dark radiation with
the already existing constraints from searches for new neutral gauge
bosons at LHC7.
\end{abstract}
\end{frontmatter}

Additional heavy neutral $Z'$ gauge bosons are predicted in many
extensions of the standard model (SM) such as most Grand Unified Theories
(GUT) and superstrings \cite{Langacker:2008yv}. 
Generically a model with an extra $U(1)'$ gauge symmetry 
is characterized by the mass of the $Z'$, the $U(1)'$ gauge coupling and
the chiral charges of the matter fields (in some cases including
additional particles with exotic SM charges to cancel anomalies), 
and the possible mixing angle  between the $Z$ and $Z'$. 
Particularly well motivated and consistently anomaly-free constructions 
are $Z'$ models based on the E$_6$ GUT group and for this reason they have 
been extensively  studied in the literature. In what follows we will 
concentrate on these type of models though the results here presented
can be easily generalized to other assignments of $Z'$ charges and couplings.

Since  E$_6$  is a rank 6 group, it contains in general two neutral
gauge bosons beyond those of the SM. These couple to two new hypercharges
$\psi$ and $\chi$ corresponding to the $U(1)$ symmetries in E$_6/{\rm SO(10)}$ 
and ${\rm SO(10)}/{\rm SU(5)}$ respectively. These hypercharge quantum 
numbers for the SM fermions  and
right-handed neutrinos are given in Table \ref{tab:charges}.
\begin{table}
\begin{center}
\begin{tabular}{|c|c|c|c|c|}
\hline
&$T_3$& $Y$ & $\sqrt{40}Y_\chi$ & $\sqrt{24}Y_\psi$ \\
\hline
$Q_=L\left(\begin{array}{c}u_L\\[-0.2cm]d_L\end{array}\right)$ 
&$\left(\begin{array}{c}1/2\\[-0.2cm]-1/2\end{array}\right)$ 
& $1/6$ & $-1$ & $1$ \\
\hline
$u_R$ & 0 & $2/3$ &   $1$ & $-1$ \\
\hline
$d_R$ & 0 & $-1/3$ & $-3$ & $-1$ \\
\hline
$L_L=\left(\begin{array}{c}\nu_L \\[-0.2cm]e_L\end{array}\right)$ 
& 
$\left(\begin{array}{c}1/2\\[-0.2cm]-1/2\end{array}\right)$ & $-1/2$
&  $3$ & $1$ \\
\hline
$e_R$ & 0 & -1 &  $1$ & $-1$ \\
\hline
$\nu_R$ & 0& 0 & $5$ & $-1$ \\
\hline
\end{tabular}
\end{center}
\caption{The SM quantum numbers and hypercharges of the $U(1)_\chi$ and 
the $U(1)_\psi$ of the relevant fields.}
\label{tab:charges}
\end{table}
Here we focus on the case where the gauge sector contains only one
additional $U(1)$ symmetry at low energies. Therefore there is a
continuum of possible models where the new gauge boson couples to 
one linear combination of $Y_\chi$ and $Y_\psi$ 
parametrized by a mixing angle $\beta$ 
\footnote{A special case that is often considered in the literature 
is $U(1)_\eta$, which in our convention corresponds to 
$\beta =\tan^{-1}(-\sqrt{5/3})\simeq 0.71 \pi$}.
\begin{equation}
Y_\beta=\cos\beta\, Y_\chi + \sin\beta \, Y_\psi \; , 
\end {equation}
In what follows will make our study for an arbitrary value of $\beta$
which corresponds to the most general single $Z'$ model where the new
$U(1)$ can be embedded in a primordial E$_6$ symmetry.  We can chose
$0\leq \beta\leq \pi$ since the charges merely change sign for $\beta
\rightarrow \beta + \pi$. In this case the sign of the mixing angle
between $Z$ and $Z'$ becomes physical. However the latest analysis of
precision electroweak data \cite{Erler:2009jh} constraints the
$Z$--$Z'$ mixing angle to be at most ${\mathcal O}(10^{-3})$ 
for $M_{Z'}\lesssim 1$ TeV and theoretically
the mixing angle is expected to decrease inversely proportional to $M_{Z'}^2$.
For the sake of simplicity we will neglect the small effects associated 
with  the $Z$--$Z'$ mixing and set it to zero in our calculations.  
Finally the value of the 
$U(1)_\beta$  coupling 
is fixed by the condition of coupling constant unification 
$g_{Y_\beta}=\sqrt{\frac{5}{3}}g_2$  where $g_2$ is the $SU(2)$ coupling
constant.  

\begin{figure}
\centering\includegraphics[width=0.7\textwidth]{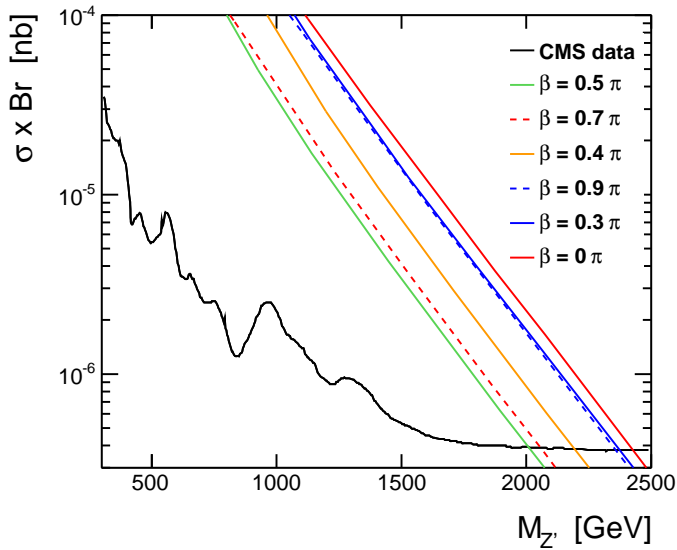}
\caption{95\%  upper limits on the productino ratio 
$\sigma(pp \rightarrow Z') \times {\rm Br}(Z'\rightarrow l^+l)$
from the combined di-muon and di-electron searches in Ref.\cite{CMS}
as a function of the resonance mass. Also shown in the figure the
predictions for E$_6$ $Z'$ models for different values of 
the model parameter $\beta$ as labeled in the figure. 
The curves corresponds from left to right 
to the values of $\beta$ listed from top to bottom.}
\label{fig:lhcbounds}
\end{figure}

\begin{figure}
\centering\includegraphics[width=1\textwidth]{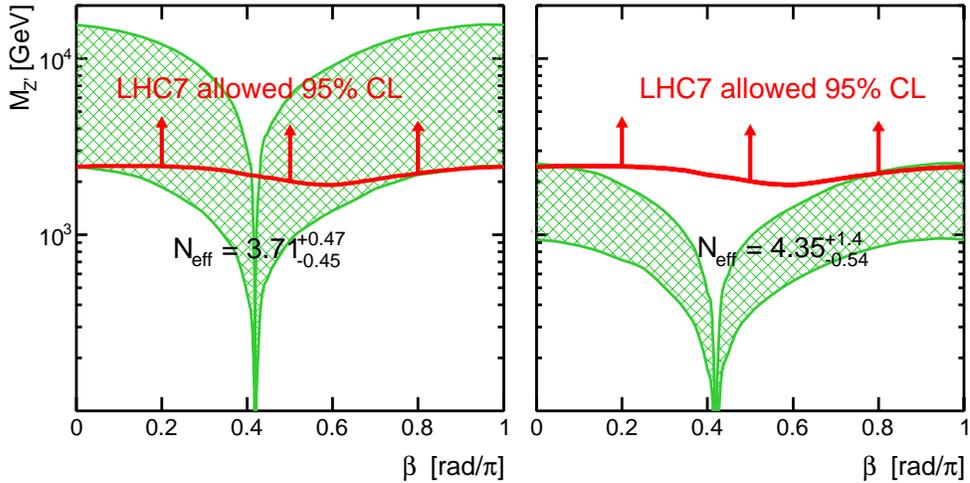}
\caption{Required  values of $M_{Z'}$ for two illustrative 
1$\sigma$ ranges of $N_{\rm eff}$, $N_{\rm eff} =4.35^{+1.4}_{-0.54}$  
($0.76\leq\Delta N_{\rm eff}\leq 2.7$)
as obtained in the analysis of
Ref.~\cite{GonzalezGarcia:2010un}  (left panel) 
and for the lower range presently
favoured by BBN nucleosynthesis $N_{\rm eff} =3.71^{+0.47}_{-0.45}$
($0.21\leq\Delta N_{\rm eff}\leq 1.13$)
(right panel). The solid red curve are the 
95\% CL lower bounds on $M_{Z'}$ from  searches for $Z'$ resonance 
in di-lepton  production at LHC4 in Ref.\cite{CMS}.}
\label{fig:mzbeta}
\end{figure}

If present at the TeV scale these $Z'$ bosons would lead to
unmistakable signatures at colliders with the strongest constraints
arising from searches of resonances decaying into dilepton final
states at LHC7 ~\cite{CMS,ATLAS}. In Figure \ref{fig:lhcbounds} we show
the strongest present bounds on $\sigma(pp \rightarrow Z') \times {\rm
Br}(Z'\rightarrow l^+l)$ at 95\% CL from these searches obtained with
5fb$^{-1}$ luminosity \cite{CMS} together with the theoretical
expectations in the E$_6$ $Z'$ models for different values of
$\beta$. The theoretical expectations were obtained introducing
the E$_6$ models in the package MADEVENT~\cite{madevent} and with the
CTEQ6L parton distribution functions \cite{CTEQ6}. They are in perfect
agreement with the corresponding theoretical curves shown by the
collaborations for some specific values of $\beta$. From the figure
we read the most up-to-date bound on $M_{Z'}$ as a function of the model 
parameter $\beta$ which we will use in the following and it is shown 
in Fig.\ref{fig:mzbeta}.

A TeV-scale $Z'$ also has important cosmological
implications.  Since the right-handed neutrinos carry a non-zero $U(1)'$
charge, the $U(1)'$ gauge symmetry  prevents the large
right-handed Majorana masses needed for the ordinary neutrino seesaw
mechanism. In this case right-handed neutrinos remain massless while
the left-handed SM states must be light to account for the oscillation
data \cite{GonzalezGarcia:2007ib}. Alternatively right-handed and
left-handed neutrinos can combine to form three light Dirac neutrinos.
Either way the lightness of the observed neutrino masses is generated
by some additional mechanism and the three right-handed states are
either light or massless and only couple to the SM sector via the $Z'$
interactions. In this case the theory contains the right-handed neutrinos 
relativistic degrees of freedom in addition to photons and the 
three left-handed neutrinos of the SM which can contribute to 
the expansion rate of the Universe as a new form of ''dark radiation'' 
and consequently affect the cosmological observations. 

In particular faster expansion would lead to an earlier freeze-out of
the neutron to proton ratio and would lead to a higher $^4$He
abundance generated during Big Bang Nucleosynthesis (BBN)
\cite{Yang:1978ge}. At later times dark radiation would also alter the 
time for the matter-radiation equality with the corresponding impact 
in the observed cosmic microwave background (CMB) anysotropies as well 
as affect the large scale structure  (LSS) distributions . 

\begin{figure}
\centering\includegraphics[width=0.49\textwidth]{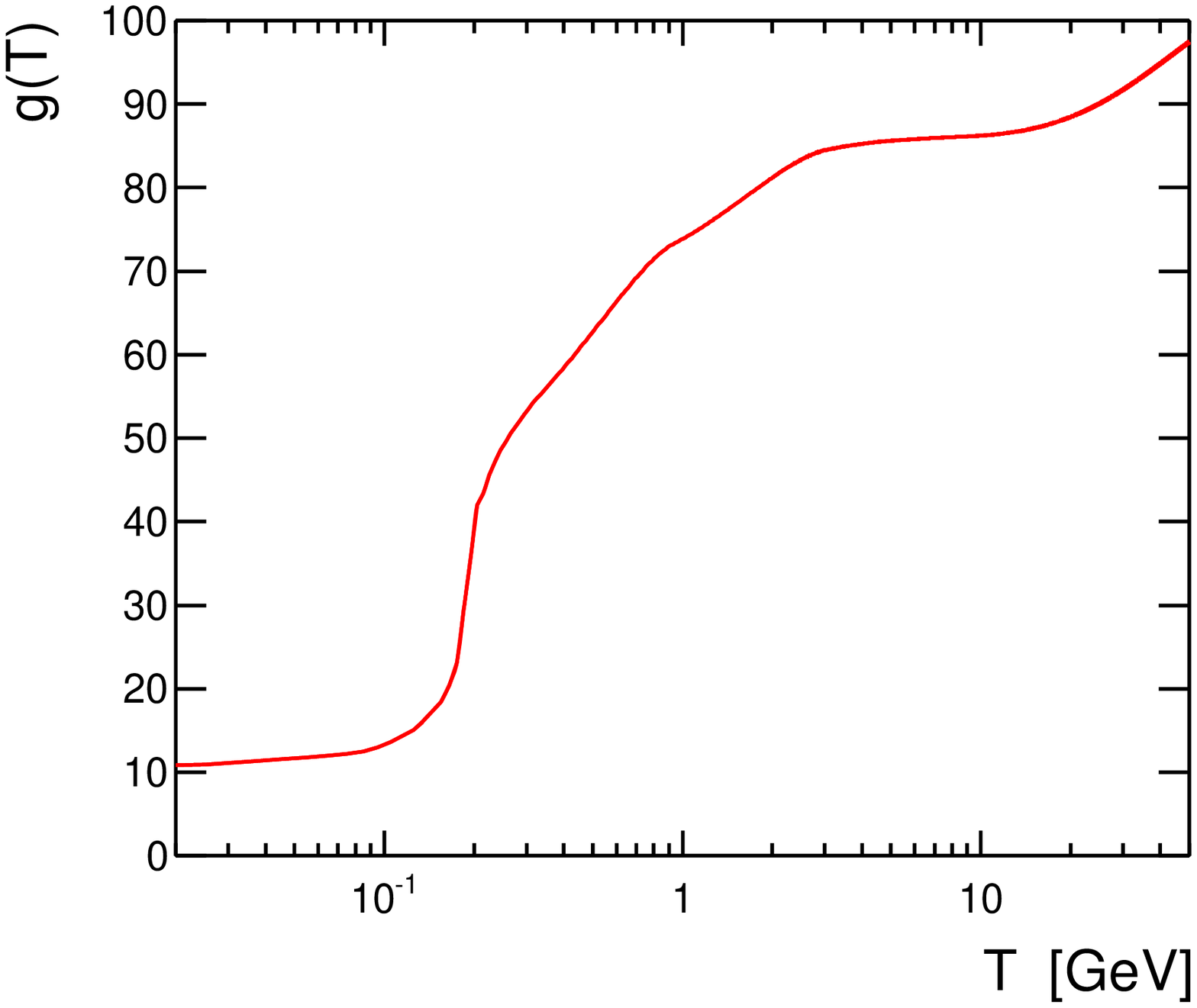}
\centering\includegraphics[width=0.49\textwidth]{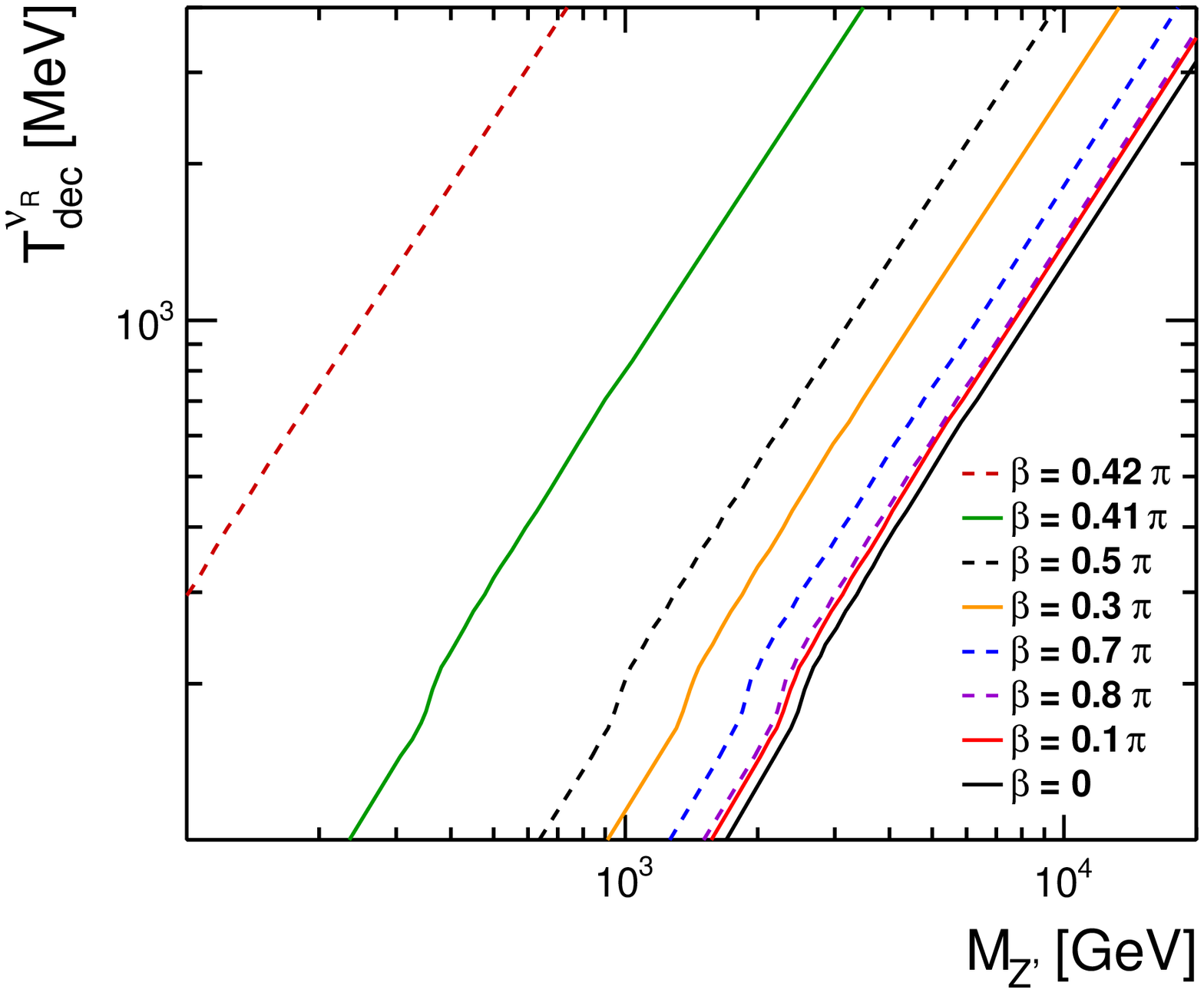}
\caption{Left:The effective number of degrees of freedom as a function
of the temperature without including the contribution of the
right-handed neutrinos. Right: Decoupling temperature of the right-handed
neutrinos as a function of the $Z'$ mass for different values of
the model parameter $\beta$ as labeled in the figure.
The curves corresponds from left to right 
to the values of $\beta$ listed from top to bottom.}
\label{fig:gt-tdec}
\end{figure}

In Refs.~\cite{Steigman:1979xp,Olive:1980wz} it was first discussed in
the context of BBN the implications of a superweakly interacting light
particle, such as the right-handed neutrinos coupling to a heavy
$Z'$. Because of their superweak interactions, such particles decouple
earlier than ordinary neutrinos and consequently their contribution to
the energy density budget of the Universe -- and therefore to its
expansion -- is suppressed with respect to that of the left-handed
neutrinos. 
The cosmic radiation content is usually expressed in terms of the
effective number of thermally excited neutrino species, $N_{\rm eff}$
with its standard value being $N_{\rm eff}=3.046$. After their 
decoupling (for $T<T^{\nu_L}_{\rm dec}<T^{\nu_R}_{\rm dec}$) the  
three superweakly interacting light right-handed neutrinos  
contribute to  $\Delta N_{\rm eff}$ as: 
\begin{equation}
\Delta N_{\rm eff}=3\times \left(\frac{T_{\nu_R}}{T_{\nu_L}} 
\right)^4=3\times \left(\frac{g(T_{\rm dec}^{\nu_L})}
{g(T^{\nu_R}_{\rm dec})}\right)^\frac{4}{3} 
\label{eq:neff}
\end{equation}
where $g(T)$ is the effective number of degrees of freedom 
at temperature $T$. 
Neglecting finite mass corrections,  $g(T)=g_B(T) +
\frac{7}{8} g_F(T)$, where $g_{B,F}(T)$ are the number of bosonic and
fermionic relativistic degrees of freedom in equilibrium at
temperature $T$. Thus at $T^{\nu_L}_{\rm dec}\sim $ 3 MeV, 
$g(T_{\rm dec}^{\nu_L})=43/4$ corresponding to three active neutrinos,
$e^\pm$ and photons. 
In calculating $g(T)$ at higher temperatures
one must take into account the QCD phase transition  
at temperature $T_c$. Above $T_c$ the 
quarks and the gluons are the relevant hadronic degrees of freedom, 
while below $T_c$ they are replaced by the hadronic degrees of freedom.
At present the  estimated value of  $T_c=154 \pm 9$ MeV, 
\cite{Bazavov:2011nk} and the evolution of the energy and entropy 
across the QCD phase transition are  obtained by means of 
state of the art lattice QCD simulations \cite{Bazavov:2009zn}. 
Using the results in Ref. \cite{Bazavov:2009zn} and including finite
mass effects we obtain the $T$ 
dependence of $g(T)$ (without the right-handed neutrino contribution) 
shown in the left panel in Fig.~\ref{fig:gt-tdec}. 

In the $Z'$ models here discussed right-handed neutrinos are kept in 
equilibrium by their interactions with the SM fermions mediated by 
the $Z'$  so
\begin{eqnarray}
\Gamma_{\nu_R}(T)&=&{\displaystyle \sum_f}\Gamma_{f}(T)
={\displaystyle \sum_f} \frac{g_{\nu_{R}}}{n_{\nu_{R}}}
\langle v \sigma(\bar\nu_R\nu_R \rightarrow \bar f f)\rangle
\nonumber \\
&\equiv &
{\displaystyle \sum_f} \frac{g_{\nu_{R}}}{n_{\nu_{R}}}
\int \frac{d^3p}{(2\pi)^3} \frac{d^3 q}{(2\pi)^3} f_{\nu_R}(q) f_{\nu_R}(p) 
\sigma_f(s) \, v
\label{eq:intrate} \\
&=& 
{\displaystyle \sum_f} 
\frac{g_{\nu_{R}}}{8\pi^{4}n_{\nu_{R}}}\ \int_0^\infty
p^2\,dp\ \int_0^\infty q^2\,dq \int_{-1}^{1}
\frac{(1-\cos\theta)\sigma_{f}(s)}{(e^{q/T}+1)(e^{p/T}+1)}\,d\cos\theta \;. 
\nonumber
\end{eqnarray}
$g_{\nu_R}$=2 and  $n_{\nu_{R}}$ is the number density of a 
single generation of right-handed neutrinos which follows a 
Fermi-Dirac distribution $f_{\nu_R}(k)=\frac{1}{e^k/T+1}$. 
$s=(p+q)^2=2pk(1-\cos\theta)$ is the COM energy in the collision and
$v=(1-\cos\theta)$ where $\theta$ is the relative angle of the
colliding right-handed neutrinos. 

In Eq.~(\ref{eq:intrate}) the annihilation cross section is
\begin{eqnarray}
\sigma_f(s)&\equiv&
\sigma(\bar{\nu_{R}}\nu_{R}\rightarrow\bar f f )  \\
&=&
N_{C}^{f}\frac{s\beta_f}{16\pi} \frac{g_{Y_\beta}^2}{M^{2}_{Z'}}
\left(Y_\beta^{\nu_{R}}\right)^2 
\left\{(1+\frac{\beta_{i}^{2}}{3})\left[
\left(Y_\beta^{f_{L}}\right)^2 +\left(Y_\beta^{f_{R}}\right)^2 \right] 
+2(1-\beta_{i}^{2}) Y_\beta^{f_{L}}Y_\beta^{f_{R}} 
\right\}\nonumber
\end{eqnarray}
where $N_C^f$ is the number of colours of the fermion $f$ and
$\beta_f=\sqrt{1-\frac{4 m_f^2}{s}}$. 

The $\nu_R$'s will  decouple when  their interaction rate 
drops below the expansion  rate of the  Universe $H(T)$. 
Right before decoupling  
\begin{equation}
H(T)=\sqrt{\frac{4\pi^{3}G_{N}\left(g(T)+\frac{21}{4}\right)}{45}} T^2
\label{eq:HT}
\end{equation}
where we have included the $\frac{21}{4}$ contribution to the expansion
due to the 3 generations of massless $\nu_R's$ . 
The decoupling temperature is defined by the condition 
$\Gamma_{\nu_R}(T^{\nu_R}_{\rm dec})=H(T^{\nu_R}_{\rm dec})$. Its value
as a function of $M_{Z'}$ and $\beta$  can be found 
by numerically solving this equality with the expressions in 
Eqs.~(\ref{eq:intrate}) and~(\ref{eq:HT}) and $g(T)$ as given in 
Fig.\ref{fig:gt-tdec}. We show it in the  
right panel of Fig.\ref{fig:gt-tdec}. 
The coupling  $Y^{\nu_R}_\beta$ vanishes for 
$\tan\beta=\sqrt{15}$, ie $\beta=0.4196 \pi$. Thus for this 
value of $\beta$ the $\nu_R$'s are never in equilibrium with the SM particles. 
Consequently, as seen in the figure,  
$T^{\nu_R}_{\rm dec}\rightarrow \infty$ independently of $M_{Z'}$
as $\beta\rightarrow \tan^{-1}(\sqrt{15})$. 
Conversely  $Y^{\nu_R}_\beta$  is maximum
for $\beta=0,\pi$ ie for the $U(1)_\chi$ hypercharge so for a given 
$M_{Z'}$  $T^{\nu_R}_{\rm dec}$ the lowest for this model.

\begin{figure}
\centering\includegraphics[width=0.7\textwidth]{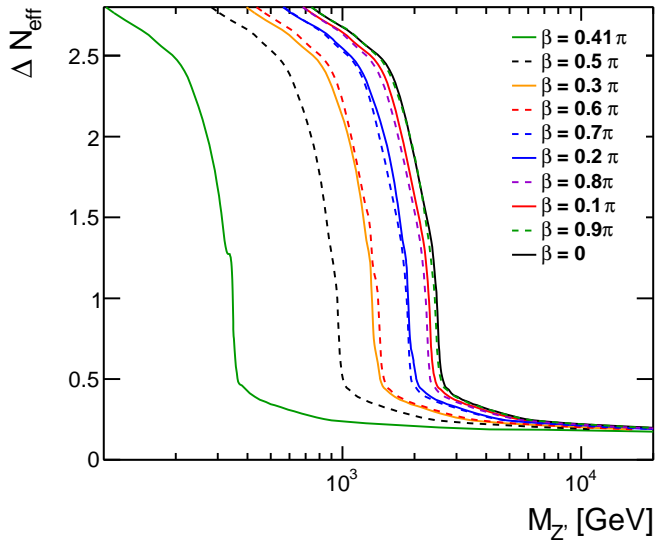}
\caption{
The equivalent number of extra neutrinos
$\Delta N_{\rm eff}$ due to 3 generation of massless $\nu_R$'s 
as a function of the $Z'$ mass for several values 
of the model parameter $\beta$ as labeled in the figure.
The curves corresponds from left to right 
to the values of $\beta$ listed from top to bottom.} 
\label{fig:neff}
\end{figure}

Once we know $T^{\nu_R}_{\rm dec}$ as a function of the model parameters
we can derive the contribution of the right-handed neutrinos to the dark 
radiation at temperatures below their decoupling as Eq.~(\ref{eq:neff}).
This is illustrated in Fig.~\ref{fig:neff} where we plot
$\Delta N_{\rm eff}$ as a function of the $Z'$ mass for several values 
of the model parameter $\beta$. As expected as $\beta$ approaches the
decoupled model $\beta=\tan^{-1}(\sqrt{15})\sim 0.42$, the additional 
number of effective neutrinos becomes small (asympotically zero) for
any value of $M_{Z'}$. 
We note that for the range of $M_{Z'}$ masses plotted in the figure,  
the minimum value of $\Delta N_{\rm eff}\simeq 0.2$ corresponds to 
the plateau around $g(T^{\nu_R}_{\rm dec})\sim 86$ 
for  the corresponding decoupling temperatures 
3 GeV$\lesssim T^{\nu_R}_{\rm dec}\lesssim 10$ GeV as can be seen 
in Fig.~\ref{fig:gt-tdec}. 

All this implies that the cosmological information on $\Delta N_{\rm eff}$ can
be used to constraint the $Z'$ properties. In particular during
most of the last two decades the measured primordial abundances of
$^4$He and other nuclei were used with this purpose
\cite{Ellis:1985fp,GonzalezGarcia:1989py,Barger:2003zh}.  Since at the
time the observed abundances were compatible with the presence of no
dark radiation, generically  {\sl lower} bounds on the $Z'$ mass were
derived.

This situation is altered at present as recent analysis of
cosmological data suggest a trend towards the existence of ``dark
radiation'' (see for example
\cite{Komatsu:2010fb,GonzalezGarcia:2010un,Hamann:2010bk,Nollett:2011aa,GonzalezMorales:2011ty,Joudaki:2012fx,Archidiacono:2012gv}).
The Wilkinson Microwave Anisotropy Probe (WMAP) collaboration found
$N_{\rm eff} = 4.34^{+0.86}_{-0.88}$ based on their 7-year data
release and additional Large Scale Structure (LSS)
data~\cite{Komatsu:2010fb} at $1 \sigma$ in a $\Lambda$CDM cosmology.
In Ref. ~\cite{GonzalezGarcia:2010un} $N_{\rm eff} =
4.35^{+1.4}_{-0.54}$ was found in a global analysis including the data
from cosmic microwave background (CMB) experiments (in particular the
from WMAP-7), the Hubble constant H0 measurement \cite{Riess:2009pu},
the high-redshift Type-I supernovae\cite{Hicken:2009df} and the LSS
results from the Sloan Digital Sky Survey (SDSS) data release 7 (DR7)
halo power spectrum~\cite{Reid:2009xm} in generalized cosmologies
which depart from $\Lambda$CDM models by allowing not only the
presence of dark radiation but also dark energy with equation of state
with $\omega\neq -1$, neutrino masses, and non-vanishing
curvature. More recent measurements of the CMB anisotropy on smaller
scales by the Atacama Cosmology Telescope (ACT) \cite{Das:2010ga} and
South Pole Telescope \cite{Keisler:2011aw} experiments seem to also
favour a value of $N_{\rm eff}$ higher than predicted in SM.
Cosmological constraints from BBN indicate, as well, that the
relatively high $^4$He abundance can be interpreted in terms of
additional radiation during the BBN
epoch~\cite{Izotov:2010ca,Aver:2010wq,Mangano:2011ar} (see
\cite{Steigman} for a recent review).  In Ref.\cite{Steigman} using
observed D and $^4$He abundances and fitting simultaneously $N_{\rm
eff}$ and the the baryon asymmetry, it is found $N_{\rm eff}=
3.71^{+0.47}_{-0.45}$ at 1$\sigma$.

This positive evidence of a non-zero $\Delta N_{\rm eff}$ can be
interpreted in the $Z'$ models as evidence of a TeV scale $Z'$ and
used to quantitatively derive the required values of $M_{Z'}$ as a
function of the model parameter $\beta$ and to compare those with the 
present LHC7 constraints \footnote{Along these
lines Ref.\cite{Anchordoqui:2011nh} presents an estimate of the
corresponding gauge boson mass range for an specific model with two
additional $Z'$ bosons.  See also Ref.\cite{Anchordoqui:2012wt}.}.  
We find that the 95\%CL LHC7  lower bounds on  $M_{Z'}$ as derived  from 
Fig.~\ref{fig:lhcbounds} imply that within these scenarios the effective
number of neutrinos is bounded to be \footnote{In particular for the
$\eta$ model $\Delta N^{max}_{\rm eff}=0.42$.}
\begin{equation}
\begin{tabular}{|@{\extracolsep{0.1cm}}c|@{\extracolsep{0.1cm}}
c|@{\extracolsep{0.1cm}}c
|@{\extracolsep{0.1cm}}c|@{\extracolsep{0.1cm}}c|@{\extracolsep{0.1cm}}c|@{\extracolsep{0.1cm}}c|@{\extracolsep{0.1cm}}c|@{\extracolsep{0.1cm}}c|@{\extracolsep{0.1cm}}c|@{\extracolsep{0.1cm}}c|@{\extracolsep{0.1cm}}c|}
\hline
$\Delta N^{max}_{\rm eff}$& 1.16 &0.48 &0.37 &0.30 &0.22 &0 &0.27 &0.36 
&0.43 &0.98 &1.18 \\\hline
for $\frac{\beta}{\pi}$ &0.00 &0.10 &0.20 &0.30 &0.40 &0.419& 0.50 &0.60& 
0.70& 0.80& 0.90 \\\hline
\end{tabular}
\label{eq:neffmax}
\end{equation}

We also show the results of this exercise in Fig.\ref{fig:mzbeta}
where we plot the values of $M_{Z'}$ required to generate two
illustrative 1$\sigma$ ranges of $N_{\rm eff}$, $N_{\rm eff}
=4.35^{+1.4}_{-0.54}$ ($0.76\leq\Delta N_{\rm eff}\leq 2.7$) as
obtained in the analysis of Ref.~\cite{GonzalezGarcia:2010un} (right
panel) and for the lower range presently favoured by BBN
nucleosynthesis $N_{\rm eff} =3.71^{+0.47}_{-0.45}$ ($0.21\leq\Delta
N_{\rm eff}\leq 1.13$) (left panel). Also shown in the figure are 95\%
CL the present bound from LHC7 as obtained from
Fig.~\ref{fig:lhcbounds}. As seen in the figure already with the
existing bounds from LHC7 most of the values of $M_Z'$ required to
account for the larger amount of dark radiation 
($N_{\rm eff} =4.35^{+1.4}_{-0.54}$) are disfavoured. 

In summary in this letter we have explored the possibility of accounting 
for the increasing evidence of dark radiation found in recent analysis 
of CMB, LSS, and BBN data in the framework of E$_6$ $Z'$ models in the 
light of  the present LHC7 constraints  on these scenarios 
(which we display in Fig.\ref{fig:lhcbounds}).
In these models additional radiation exists in the form 
of three generations of relativistic right-handed neutrinos in an amount
that depends on how long they are kept in equilibrium with the SM particles 
by the $Z'$ interactions as we quantify in 
Figs.~\ref{fig:gt-tdec} and~\ref{fig:neff}. 
We conclude that within the present 95\% CL bounds already imposed by 
LHC7, these scenarios cannot account for extra radiation in excess of 
1.25 effective neutrino species for any value of $\beta$ and in excess of 
0.5 effective neutrinos for models with $0.1<\beta/\pi<0.75$ 
(see Eq.~(\ref{eq:neff})). 
The potential of the future LHC14 runs to further constraints
the amount of dark radiation in these scenarios can be easily read 
from Fig.\ref{fig:neff}. For example, should no resonance be found 
at LHC14, it is foreseeable that the lower bound 95\% of $M_{Z'}$ 
will become  at least $\sim$ 5 TeV~\cite{LHC14}. In this case  
the amount of  dark radiation in these $Z'$ models will be limited to 
at most  that of 0.3 effective neutrinos.

\vskip 1cm
We thank C. Manuel and K. Rajagopal for clarifications and references 
about lattice simulations of the QCD phase transition. We also thank J. 
Taron for careful reading of the manuscript and F. Dias for help with 
references for the LHC14 sensitivites.
A.Solaguren-Beascoa. thanks J. Gonzalez-Fraile and F. Mescia for techinical help with using MADEVENT.
This work is supported by Spanish MICINN grant FPA2010-20807,
and consolider-ingenio 2010 grant
CSD-2008-0037, by CUR Generalitat de Catalunya grant 2009SGR502, 
by USA-NSF grant PHY-0653342 and by EU grant 
FP7 ITN INVISIBLES (Marie Curie Actions PITN-GA-2011-289442).

\end{document}